%% file: AAuthorReview.tex
\documentclass[acmsmall]{acmart} 
\input{premeable}
\PassOptionsToPackage{prologue,table,xcdraw}{xcolor}
\newcommand{\one}[1]{\textcolor{black}{#1}}
\begin{document}

\title{From Regulation to Support: Centering Humans in Technology-Mediated Emotion Intervention in Care Contexts}

\author{Jiaying "Lizzy" Liu}
\affiliation{%
  \institution{School of Information, The University of Texas at Austin}
  \country{USA}}
\email{jiayingliu@utexas.edu}

\author{Shuer Zhuo}
\affiliation{%
  \institution{Advertising, The University of Texas at Austin}
  \country{USA}}
  \email{zhuoshuer@utexas.edu}

\author{Xingyu Li}
\affiliation{%
  \institution{Digital Media, Georgia Institute of Technology}
  \country{USA}}
\email{xingyu@gatech.edu}

\author{Andrew Dillon}
\affiliation{%
  \institution{School of Information, University of Texas at Austin}
  \country{USA}}
\email{adillon@ischool.utexas.edu}

\author{Noura Howell}
\affiliation{%
  \institution{Digital Media, Georgia Institute of Technology}
  \country{USA}}
\email{nhowell8@gatech.edu}

\author{Angela D. R. Smith}
\affiliation{%
  \institution{School of Information, The University of Texas at Austin}
  \country{USA}}
\email{angela.smith@ischool.utexas.edu}

\author{Yan Zhang}
\affiliation{%
  \institution{School of Information, The University of Texas at Austin}
  \city{Austin}
  \country{USA}
}
\email{yanz@utexas.edu}

\renewcommand{\shortauthors}{Liu et al.}

\begin{abstract}
Enhancing emotional well-being has become a significant focus in HCI and CSCW, with technologies increasingly designed to track, visualize, and manage emotions. However, these approaches have faced criticism for potentially suppressing certain emotional experiences. Through a scoping review of 53 empirical studies from ACM proceedings implementing Technology-Mediated Emotion Intervention (TMEI), we critically examine current practices through lenses drawn from HCI critical theories.
Our analysis reveals emotion intervention mechanisms that extend beyond traditional emotion regulation paradigms, identifying care-centered goals that prioritize non-judgmental emotional support and preserve users' identities. 
The findings demonstrate how researchers design technologies for generating artificial care, intervening in power dynamics, and nudging behavioral changes. We contribute the concept of "emotion support" as an alternative approach to "emotion regulation," emphasizing human-centered approaches to emotional well-being. This work advances the understanding of diverse human emotional needs beyond individual and cognitive perspectives, offering design implications that critically reimagine how technologies can honor emotional complexity, preserve human agency, and transform power dynamics in care contexts.
\end{abstract}

\maketitle

\section{Introduction}

\one{
Technology-mediated emotion interventions (TMEIs) encompass digital technologies engineered to recognize, analyze, and modulate human emotional states. Some common examples include biometric wearables that monitor individuals' physiological indicators of stress \cite{saganowski_emotion_2023} and conversational agents designed to offer empathetic responses \cite{10.1145/3290605.3300932}. 
These interventions have proliferated in recent years across workplace settings \cite{thomas_mental_2022} and daily life \cite{m_smartphone_2018}. In particular, in care contexts, TMEIs are increasingly designed to improve the emotional well-being of individuals with various health conditions \cite{huh_health_2014} and nurture vulnerable groups, such as children and older adults \cite{theofanopoulou_exploring_2022}.
}

\one{
However, scholars raise significant accuracy and ethical concerns regarding TMEIs that label and modify human emotions \cite{stark_ethics_2021}. Researchers, such as \textcite{andalibi_human_2020}, argue that emotion recognition algorithms fundamentally fail to account for the subjective and contextual nature of emotions while imposing rigid categorizations that cannot capture how emotional experiences vary across individuals and cultural contexts \cite{10.1145/3442188.3445939}.
Moreover, TMEIs often medicalize and attempt to "fix" emotional states, problematically prioritizing the "correction" of detected emotions \cite{pendse_treatment_2022}. These systems frequently prescribe simplistic interventions, such as prompting users to "cheer up" when sad or offering breathing exercises to eliminate anger. 
This normative approach may impose unreasonable expectations about "proper" emotional expression, which can disproportionately harm vulnerable populations in care contexts. People with health conditions may have entirely appropriate emotional responses to their circumstances that are wrongly flagged as problematic. Similarly, children and older adults—who often have less power in care settings—may find their legitimate emotions dismissed or "corrected" according to standardized norms that do not account for their unique situations \cite{10.1145/3544548.3581480}.
In this vein, critics further contend that these digital tools may function as vectors for colonialism and oppression \cite{chordia_social_2024}, creating systems of emotional surveillance through monitoring and tracking technologies that particularly impact those with the least agency in care relationships \cite{choe_emotion_2023}.
}

\one{
In response to these concerns, a growing body of work in HCI has advocated for justice- and ethics-oriented research \cite{chordia_social_2024}, deliberately recentering the intentions and needs of the people that TMEIs aim to support. Scholars have called for design practices that move beyond a categorical understanding of emotions as biologically determined and universally expressed \cite{kitson_supporting_2024, petr_slovak_designing_2023}.
Researchers are exploring technologies that honor individuals' lived experiences through approaches that recognize emotions as co-constructed within social contexts \cite{10.1145/3610073, 10.1145/3512939} and embodied through physical sensations \cite{10.1145/3064663.3064697, 10.1145/3491102.3502135, 10.1145/3490149.3501326}. For example, Lazar et al. \cite{10.1145/3359187} radically support people with dementia by curating a non-judgmental online community space where participants can safely share and communicate their "real and raw" emotional experiences without fear of stigma.
Similarly, researchers have worked to broaden our understanding of emotional needs by incorporating perspectives from historically underrepresented populations. Bhattacharjee et al. \cite{bhattacharjee_whats_2023} documented how Indian people often avoid helpline-delivered talk therapy rooted in Western psychology, highlighting the cultural specificity of emotional interventions and the need for culturally responsive approaches.
}

\one{
Joining this line of work, our study recognizes the potential harm that digital tools can inflict on human feelings \cite{pendse_treatment_2022}, identities \cite{pendse_marginalization_2023}, and lived experiences \cite{feuston_everyday_2019}. We deliberately employ critical theories as reflexive lenses to foreground and interrogate the power dynamics embedded within TMEIs \cite{hooks_feminism_2000}. Our research asks:
\begin{itemize}
    \item RQ1: What approaches do TMEIs adopt in modulating the emotional experiences of people?
    \item RQ2: What goals do TMEIs establish as the ideal outcomes of emotion interventions?
    \item RQ3: What roles do technologies play in the care ecosystem when implementing TMEIs?
\end{itemize}
}

\one{
We conducted a scoping literature review of the ACM Digital Library, investigating the design of 53 empirical studies that implemented technology-mediated emotion interventions (TMEIs). 
Our analysis identified specific approaches (RQ1) through which technology intervenes in human emotions, revealing approaches that extend beyond narrowly adopting psychological theories around emotion regulation \cite{gross_antecedent-_1998} and cognitive behavioral therapy \cite{beck201960}. Furthermore, we mapped the goals of TMEI tools (RQ2) to untangle how they structure and influence users' understanding of and interaction with their feelings. 
Our findings revealed a notable intervention dichotomy: technologies predominantly aimed to down-regulate negative emotions while up-regulating positive affect and motivation. We also identified additional care-centered goals, such as non-judgmental approaches and identity preservation.
Recognizing emotions as socially constructed, we situated these technological interventions within the broader care ecosystem encompassing individuals requiring care, informal caregivers, health professionals, and institutions. This contextual analysis uncovered three prominent technological roles (RQ3): generating artificial care, intervening in power dynamics, and nudging behavioral changes.
}

\one{
These findings underscore the need to recenter humans in technology-mediated emotion interventions (TMEIs), moving beyond current HCI practices that simply replicate the emotion regulation paradigm from psychology. We introduce "emotional support" as a core value informed by critical HCI scholarship that can guide future TMEI design and implementation.
This study emphasizes the importance of contextual, culturally sensitive, and empowering approaches to emotional well-being in technological interventions. Specifically, we make the following contributions: 
\begin{itemize}
\item We critically examine technology-mediated emotion intervention practices and propose a human-centered "emotion support" approach that validates humans' emotional experiences, acknowledging the complex, socially constructed nature of emotions.
\item We delineate the roles technologies play within the caregiving ecosystem. Our analysis reveals the potential of these technologies to mitigate caregivers' emotional labor, preserve individuals' identities throughout their care-seeking journeys, and recalibrate power dynamics between caregivers and care receivers. 
\end{itemize}
}

\section{Related Work}
In this section, we review the emerging reflection on technology-mediated emotion interventions (TMEIs) in HCI and CSCW and center on the frictions of deploying such computational and technology designs in care contexts. Further, we incorporate HCI critical scholarships to attend to the power dynamics among technologies and humans in the caregiving process, reflecting on the current and future roles of technologies. 


\subsection{Potential Harms of Technology-Mediated Emotion Intervention}
\one{
Recent HCI studies have begun to examine the limitations of current practices in areas such as Affective Health and Digital Mental Health. Many studies commonly analyze users' emotional states through various inputs—sensory data (e.g., heart rates, skin temperature), facial expressions, and social media content \cite{james1948emotion}. These classifications often build upon Paul Ekman's influential work \cite{ekman1992there}, which initially proposed six basic emotions (anger, disgust, fear, joy, sadness, and surprise).
While Ekman's framework was later expanded, it has been critiqued for potentially reinforcing the questionable notion that emotions exist as distinct, separable phenomena, thereby oversimplifying the complex, dynamic, and fluid nature of human emotional experiences \cite{barrett_experience_2007}. Rather than risking a reductive approach that labels users' emotional states according to rigid classifications, some researchers call for a more nuanced understanding of emotions \cite{10.1145/3637383} beyond merely computationalizing people's emotional states.
\textcite{roemmich_emotion_2024}, in their review of emotion recognition algorithms, suggest that "\textit{emotions are constructed by individuals within social contexts, shaped by personal and socio-cultural differences rather than universally expressed; emotional experiences are structured from multiple overlapping, fundamental affective dimensions instead of discrete categories}."
}

The classification of emotions may be further complicated by algorithm errors and biases. While developers typically prioritize the algorithmic accuracy of emotion recognition, this technical focus often overlooks critical user concerns. Many users find such precise emotional inferences uncomfortable and potentially threatening to their agency, raising significant privacy concerns \cite{grill_attitudes_2022}. This discomfort stems from the invasive nature of emotion recognition technologies, which users frequently associate with potential loss of autonomy and control \cite{andalibi_human_2020, mcstay2020emotional}. For example, applying emotion recognition in the workplace may create additional emotional labor for workers to align with organizational expectations \cite{roemmich2023emotion}, potentially impacting a worker's autonomy. However, the mistakes of algorithms can be difficult to detect \cite{corvite_data_2023}, and such labor may disproportionately affect low-power and marginalized workers, possibly contributing to broader issues of data colonialism \cite{boyd_automated_2023}. Moreover, in increasingly algorithmic-moderated online platforms, inherent algorithmic biases toward certain sentiments in content recommendation systems might inadvertently impact specific narratives \cite{feuston_everyday_2019}, potentially amplifying some voices while diminishing others and creating additional challenges for people aiming to find others with similar experiences \cite{feuston_beyond_2018}.

\one{
In line with the reflection on the harms of emotion recognition technologies, this review aims to examine the current HCI practices of technology-mediated emotion interventions and contribute to a careful design of computing and technology that does not perpetuate the oppression of human emotions and experiences.
}

\subsection{Frictions of Technology-Mediated Emotion Intervention in Care Contexts}
Beyond the limitations of current emotion algorithms, technology-mediated emotion interventions (TMEIs) encounter frictions in care contexts, such as mental health \cite{jordans_community-_2019}, chronic diseases \cite{smriti_emotion_2024}, sexually transmitted diseases \cite{singh_emotion_2019}, and vulnerable populations \cite{10.1145/3340555.3353747, 10.1145/3131607}. 

One widespread application of TMEIs involves tracking and regulating negative emotions within health management contexts, where fluctuating emotions —such as fear and anxiety \cite{pyle_lgbtq_2021}—vary with treatment progress and daily challenges \cite{10.1145/2145204.2145329}. For example, many emotion-tracking technologies try help patients and caregivers monitor distress, anxiety, or sadness, providing insights that inform supportive interventions \cite{10.1145/3449153}. These interventions often fail to maintain engagement or demonstrate lasting well-being benefits \cite{eschler_emergent_2020, kelders2012persuasive}. While many works aim to prevent negative emotions through emotional interventions, psychologists emphasize the context-dependent nature of emotions—acknowledging that feelings traditionally labeled as "negative" can be helpful or harmful depending on circumstances. Research highlights the value of "bad feelings" \cite{md_good_2019, gross_emotion_2015} and suggests that difficult emotions could be better understood and processed rather than immediately mitigated as they arise. For instance, \textcite{gross_emotion_2015} explains that “\textit{episodes of fear that lead us to avoid potentially deadly fights... and episodes of anger that propel us to fight for causes we care about}”, cogently illustrating emotions’ adaptive roles. In the care context, emotions like sadness may facilitate processing grief, while frustration may motivate patients to advocate for their needs. Rather than solely focusing on emotion regulation through mitigation, interventions could benefit from approaches that validate and support the expression of emotions \cite{bowlby2008secure}, helping individuals navigate their experiences in ways that align with their psychological and social well-being \cite{linehan1993cognitive}.


\one{
More concerning are the power imbalances perpetuated by some emotion tracking and regulating systems, reproducing and amplifying existing inequities in emotional validation and care.
While personal health information technologies are designed to support individual agency in health management, they may inadvertently shift responsibility from caregivers and professionals to patients themselves \cite{ajana2017digital, depper2017we}, "turning health care into self-care" \cite{ruckenstein_datafication_2017}.
Studies frequently report a high user burden in TMEI implementations, with cognitive and emotional demands overwhelming users during extended use \cite{10.1145/3415183, 10.1145/3610073, 10.1145/3479561, zhang_designing_2021}.
As emotional experiences become quantified and classified, individuals may lose agency in defining what their feelings mean and how they should be addressed. This shift reinforces institutional control over personal emotional data, concentrating authority within healthcare systems and formal caregivers who determine what emotional states are "problematic" and which coping strategies are "appropriate" \cite{boyd2012critical, ruckenstein_datafication_2017}.
}

Furthermore, researchers have noted that digital technologies may serve as new vectors of oppression and colonialism, perpetuating Western conceptualizations of human emotion and well being \cite{pendse_treatment_2022, zhang2025identity}.
For instance, \textcite{bhattacharjee_whats_2023} documented how Indian people frequently avoid mental health helplines due to skepticism toward the underlying psychological frameworks—particularly talk and behavioral therapy approaches rooted primarily in Western research traditions. \textcite{pendse_treatment_2022} analyzed the ways digital mental health tools can reinforce historically established Western paradigms of treatment-centered approaches to mental health. These technological implementations risk pathologizing certain emotional expressions while reinforcing colonial notions of what constitutes "normal" emotional behavior while contributing to disease stigmatization and marginalization of communities labeled as "ill."


\one{
Recognizing these tensions in applying computational and digital approaches to care-related contexts, this review aims to critically examine current practices of TMEIs, specifically analyzing how they approach human emotions, what goals these interventions establish for users, and to what extent they genuinely address human emotional needs.
}
\subsection{Critical Theory in Emotion Needs in Care Contexts}
\one{
HCI and CSCW scholars are increasingly embracing critical theories such as feminism \cite{bardzell_feminist_2010}, critical race theory \cite{ogbonnaya2020critical}, and decolonialism \cite{pendse_treatment_2022} to examine how social, cultural, and political contexts shape the design, production, and deployment of computing technologies and digital tools. Collectively, these approaches carefully examine how technologies might perpetuate systemic inequalities and power structures while advocating for marginalized voices and social justice \cite{hooks_feminism_2000, kafer2013feminist}. 
We follow \textcite{bardzell_feminist_2010} and \textcite{pendse_treatment_2022} in deliberately applying critical theories to HCI practices, which approach has generated impactful work in HCI \cite{hooks_feminism_2000}. Conceptual frameworks like Haraway's "situated knowledge" \cite{haraway_situated_2013} and Costanza-Chock's "design justice" \cite{costanza-chock_design_2020} offer approaches for more inclusive and equitable technology development, which is articularly relevant for emotion-sensitive care work. 
}

\one{
In the emotion and care contexts, this critical orientation, such as Black feminism, has contributed significantly to the discourse by revealing how emotional expressions are regulated through societal power structures that privilege certain communities while marginalizing others, particularly across intersections of race, gender, and class \cite{collins2022black, ahmed2013cultural}. These perspectives prompt designers to consider how technologies might reproduce historically and politically entrenched norms of emotions.
Specifically for TMEI technologies, critical theory-informed studies uncover the risks of Emotion AI, as researchers unpack how algorithmic systems represent and interpret human emotions \cite{dignazio_data_2023}. This is powerfully demonstrated by \textcite{buolamwini_gender_2018}'s exposure of racial and gender biases in facial emotion recognition systems that consistently misclassify Black women's emotional expressions. Building on these critiques, recent work has explored alternative design approaches that challenge traditional techno-centric interventions, such as \textcite{10.1145/3563657.3595998}'s development of VR experiences that support emotional well-being through nature-inspired sensory experiences, offering more inclusive and contextually appropriate emotional support mechanisms.
}

\one{
Critical theories also offer lenses to confront the power imbalance in care contexts. The power dynamics between caregivers and care-receivers involve multifaceted relationships of authority, dependency, and negotiated autonomy \cite{bhat_we_2023}, and technological interventions can reinforce existing hierarchies within healthcare systems and care relationships \cite{kim_how_2024}. 
For example, research by \cite{berridge_control_2022, chang_dynamic_2024} examines elder care technologies, showing how they can shift decision-making power toward caregivers, potentially diminishing older adults' agency.
Overall, the datafication involved in digital tools migrates interpretive power from individuals to "data-rich" institutions, leaving individuals "data poor" in constructing their own health narratives \cite{ruckenstein_datafication_2017, boyd2012critical}. 
These dynamics intersect with historical biases in emotional interpretation. For example, the long-standing pathologization of women’s emotional expressions as “hysteria” \cite{showalter1985female} finds contemporary parallels in emotion recognition technologies that systematically misinterpret expressions based on gender and racial characteristics \cite{stark2019facial}.
}

\one{
Drawing on critical perspectives in HCI, our analysis explores how technologies implicitly encode particular understandings of emotions, how technologies transform existing power dynamics, what normative emotional frameworks technological interventions promote, and how these technologies mediate care relationships within TMEIs. By examining these interconnected dimensions, we illuminate how digital technologies shape both personal emotional experiences and broader care systems \cite{narvilkar2019bringing}, ultimately informing more equitable and empowering approaches to emotional support technology design.}

\section{Method}
\one{To examine how HCI studies design Technology-Mediated Emotion Interventions (TMEIs) in care contexts, we conducted a scoping review \cite{arksey2005scoping, rogers_umbrella_2024, stefanidi2023literature}. Our analysis focuses on identifying the approaches used to modulate emotional experiences, the established goals for ideal outcomes, and the distinct roles technologies play within care ecosystems. Through this investigation, we uncover how intervention approaches (RQ1) and goals (RQ2) define and structure the ways users understand and cope with their emotions, while also disentangling how technologies function (RQ3) when mediating emotional experiences in care settings.}

\subsection{Literature Search}
We conducted a literature search using the ACM Digital Library's full-text collection, selected for its comprehensive coverage of HCI research and technical innovations in computing. The search was performed in July 2024. Our search strategy employed three conceptual clusters of keywords: emotion-related terms (emotion* regulation OR support OR coping), care context terms (disease OR health OR care OR cancer OR dementia), and technology-focused terms (technology OR App* OR tool OR platform).

We chose these specific keyword clusters to capture various approaches to emotional support and intervention in the first cluster, include diverse healthcare contexts while ensuring relevance to care situations in the second cluster, and encompass the broad spectrum of technological solutions being developed in the third cluster. This structured approach ensures comprehensive coverage while maintaining a focus on the intersection of emotional support, healthcare contexts, and technological interventions.

\subsection{Literature Screening}
The initial search yielded 488 results. To ensure the depth and rigor of our analysis, we focused on full research papers, excluding extended abstracts, demonstrations, posters, and workshop papers. This decision was made because full papers typically provide more comprehensive methodological details and theoretical groundwork necessary for our analysis. After applying these criteria, 321 papers remained for detailed review. 
The three authors then started the screening process, each reviewing 107 papers by examining their titles, abstracts, and full texts. 
\one{As this study is motivated to interrogate the approaches, goals, and roles of technologies in the implementation of technology-mediated emotion interventions (TMEIs), our exclusion criteria removed studies without technological deployment or intervention.}This includes (1) theoretical frameworks, (2) qualitative studies that only ask about participants' perceptions and preferences around emotion coping, and (3) literature reviews without empirical data. We also excluded research not situated within care contexts, such as workplace and academic stress management or general emotional well-being outside of care settings.

During the screening process, we maintained detailed notes on papers that warranted further discussion. \one{Specifically, we marked relevant literature reviews and qualitative studies that interview participants about their emotion preferences. While these studies do not meet our inclusion criteria as they don't have actual TMEI implementation, they help deepen our understanding of people's emotional needs and coping strategies, thus, we incorporate many of them in the Discussion Section 5.}
The three authors met weekly to discuss any articles that required collective decision-making, resolve any uncertainties in applying the exclusion criteria, and ensure consistency in the screening process. In total, we identified 53 empirical studies focused on technology-mediated emotion coping in care contexts for our final analysis.

\subsection{A Coding Framework Informed by Critical Scholarships}
\one{
We draw on HCI critical scholarship's understanding of emotions and emotion intervention technologies, with reflections summarized in Section 2.3. Informed by these critical lenses, we center our analysis on power dynamics, norms, and the functional roles of technology in TMEIs. Power asymmetries emerge when researchers exercise authority in setting design objectives and intervention standards, while these technologies simultaneously have the capacity to transform existing hierarchies within care relationships. For each paper, we applied codes to capture the \textbf{approaches} of emotion intervention, the \textbf{goals} pursued, and the \textbf{roles} played by technology in the care ecosystem.
}
\paragraph{Approaches}
\one{
We adopted the design mechanisms for emotion regulation technologies from \textcite{petr_slovak_designing_2023}. Since few studies used the reminder-recommenders mechanism, we excluded it from our analysis. We used "feedback and tracking" to categorize real-time tailored biofeedback and implicit target feedback. We grouped "cognitive model development" and "awareness and reflection support" to annotate studies that highlight the cognitive pathways of emotion regulation.
Throughout our iterative coding process, we uncovered emotion intervention approaches that extend beyond traditional emotion regulation frameworks. These emerging approaches include encouraging emotional expression through writing and creating virtual realities, fostering social connection, and promoting self-acceptance. To validate these categories, we searched for related psychological theories, for instance, by tracing the cited literature in the reviewed studies, which allowed us to refine our definitions and coding schema. Table \ref{tab:EImechanism} presents the final categorization of these approaches, along with their definitions and theoretical foundations. These theories are discussed in relation to "emotion support" in Section 5.1.
}


\paragraph{Goals}
Recognizing that certain emotion regulation practices may reinforce stereotypes, such as the perceived irrationality of women, we critically examine the goals of technological interventions in shaping emotional experiences and expressions. This allows us to scrutinize how these goals might challenge or perpetuate existing power dynamics. We conducted open coding to capture multifaceted objectives, including user experience measures, health-related outcomes, and enhancements in communication between individuals and caregivers. By mapping these goals, we investigate how technologies aim to empower users or potentially reinforce societal biases in emotional support provision. Several open codes emerged from the analysis of goals, including research goals, targeted emotions for regulation, conceptualized 'ideal' emotional states, user experience metrics, and other outcome measures.

\paragraph{Roles}
Our approach is grounded in the understanding that emotions are socially constructed \cite{frijda2017laws}, profoundly influenced by societal norms, power structures, and gender roles \cite{boiger2012construction, jaggar2014love}. Within this context, we analyze how technologies position themselves in the care ecosystem, potentially reshaping traditional caregiving dynamics. This analysis encompasses the technology's role in relation to those seeking support, informal caregivers, healthcare professionals, and peers. By examining these roles, we critically assess how technology-mediated interventions may either disrupt or maintain established power structures in emotional support contexts. For example, some emerging codes include providing direct emotional support, assisting with physical labor, collaborating with professionals, and enhancing treatment efficacy.

\subsection{Data Analysis}
The first three authors began with independent open coding of 20 papers, establishing a baseline understanding and identifying preliminary themes as described in Section 3.3. They then divided the remaining papers, extracting relevant information and documenting detailed notes from deep reading. This approach ensured thorough attention to each paper while maximizing efficiency. The team met weekly to discuss papers that posed uncertainties, exchange annotations, and compare findings. 

Following the open coding stage, we transitioned to a more focused thematic analysis guided by Braun and Clarke's methodology \cite{braun2012thematic}. Each of the three authors took responsibility for an independent section - approaches, goals, and roles - conducting thematic coding across all included studies. This specialization allowed for deeper analysis within each domain while ensuring comprehensive coverage. 

In the iterative coding process, we first generated open codes closely tied to the papers' content. For instance, when analyzing technological approaches, initial codes included "mood tracking," "emotion journaling," and "social sharing." For emotional goals, we identified codes such as "reduce anxiety," "improve emotional awareness," and "facilitate expression." Regarding technological roles, example codes included "passive monitoring," "active intervention," and "communication mediator." 
We then grouped these granular codes into broader themes through multiple rounds of discussion and refinement. For example, several open codes related to emotion modification ("convert negative thoughts," "enhance positive affect," "minimize depressive symptoms") were consolidated into the higher-level theme of "up-regulation" as a key goal of technology-mediated emotion coping. Similarly, various approach-related codes ("mood detection" and "physiological sensing") were grouped under the theme of "emotion tracking."

To ensure analytical rigor, we maintained detailed memos documenting our coding decisions and theme development process in a shared Google Excel. 
During these analytical sessions, team members actively exchanged each others' interpretations and offered alternative perspectives. 
The iterative nature of our analysis allowed us to continually refine our understanding of how emotional support technologies operate in care contexts. As new patterns emerged, we revisited previously coded papers to ensure consistent application of our evolving analytical framework. 

\subsection{Limitations}
This review primarily focuses on empirical studies from prominent human-computer interaction venues in the ACM Digital Library, aligning with our situatedness in CSCW and HCI. While this approach allows for a focused analysis, it may limit the breadth of perspectives included. To mitigate this, we compare our findings with literature from psychological and health studies. \one{In addition, we acknowledge that while affect represents a distinct conceptual dimension \cite{james1948emotion, russell1989affect, tomkins2014affect}, our research specifically centers on emotion intervention implementations as the primary focus of analysis.} Further, in examining the included empirical studies, we extracted research goals based on the authors' self-described aims, which, while effective in capturing explicit intentions, may not fully encompass nuanced motivations. Our annotations aim to better capture these underlying intentions. It's important to note that our analysis primarily focuses on stated goals rather than actual outcomes. We defer a closer examination of the effectiveness of various technologies for emotion coping to future meta-review studies. 

\section{Findings}
To critically analyze current Technology-Mediated Emotion Interventions (TMEIs), we first mapped the emotion intervention approaches adopted by empirical studies. We then analyzed how these studies' research objectives to investigate their underlying assumptions about emotions and emotion regulation. Additionally, we examined the varied roles technologies play in providing emotional care and support. The annotations of the 53 reviewed studies can be found in Table \ref{tab:annotation}.

\subsection{Approaches of Technology-Mediated Emotion Interventions}
\one{
Table \ref{tab:EImechanism} outlines emotion intervention approaches that empirical studies implemented through TMEIs. It should be noted that these categories are not mutually exclusive, and there are cases where tools designed by the empirical studies belong to two categories. For example, studies designing toys that encourage children's emotional expression and provide compassion were annotated as expressive making and compassion \cite{10.1145/2544104, 10.1145/3421507}.
}
\input{tabs/EImechanisms}

\subsubsection{Emotion Tracking and Feedback}
Emotion feedback and reflection operate by transforming abstract feelings into observable patterns, enabling individuals to identify and understand their emotional experiences \cite{10.1145/3628516.3655820,10.1145/3329189.3329209}.
Manual tracking systems leverage active participation in emotion labeling, prompting users to recognize and categorize their feelings while creating digital records that reveal emotional trends over time \cite{10.1145/3429360.3468190}. Rather than directly intervening, these systems catalyze self-guided emotion management \cite{10.1145/3429360.3468190,10.1145/3544548.3581048} and bridge communication between individuals and caregivers through shared emotional data \cite{10.1145/3628516.3655820}.
Automated emotion sensing extends this mechanism through passive, continuous monitoring using multimodal inputs—facial expressions \cite{10.1145/3491102.3501925}, vocal patterns \cite{10.1145/3383313.3412244}, and physiological signals \cite{10.1145/3130943}—to map emotional dynamics without requiring active user engagement \cite{10.1145/3383313.3412244}. This approach particularly benefits those with motivational or cognitive barriers to self-tracking, though it introduces ethical considerations regarding emotion data collection \cite{10.1145/3442188.3445939,roemmich2023emotion}.

\subsubsection{Cognitive Development}
\label{Section:Chatbot}
Cognitive development mechanisms enable users to reshape thinking patterns to modulate emotional responses, frequently through conversational agents in our corpus. These systems operate via multiple pathways: guiding users to identify "thinking traps" and facilitating cognitive restructuring grounded in counseling strategies and therapeutic theories \textcite{10.1145/3613904.3642761, 10.1145/3429360.3468190}; supporting emotion regulation through direct text intervention and editing of written expressions \cite{10.1145/3613904.3642761}; and providing emotional support via language-based exchanges \cite{10.1145/3479561, 10.1145/3396868.3400898}. Beyond one-on-one interactions, some systems integrate into conversations between people \cite{10.1145/3173574.3173905}, creating scaffolded environments for peer-to-peer emotional communication and mutual cognitive development.

\subsubsection{Social Connection}
Social connection emerges as a powerful approach, primarily facilitated through communication support technologies. Our analysis identified mobile and web applications as key technological enablers that enhance interpersonal emotional exchanges. Research consistently demonstrates that these technologies foster emotional well-being by leveraging the contagious nature of emotion processing that naturally occurs in peer interactions \cite{guo2020preliminary}.
The effectiveness of social connection as an intervention mechanism manifests through three distinct technological approaches: First, systems that enhance emotional resonance and empathetic quality in conversations through specialized interactive features \cite{10.1145/3442381.3450097,10.1145/3610073}; second, technologies that provide contextually relevant information to guide users through emotionally charged decision-making processes \cite{10.1145/3639701.3656304}; and third, platforms that deliberately structure interaction around peer support networks and emotional sharing functionalities \cite{10.1145/3415183}. These approaches collectively harness social dynamics to create technology-mediated pathways for emotional regulation and support.

\subsubsection{Expressive Making}
Expressive making encompasses interventions that facilitate emotional expression through creative activities such as writing and artistic creation. Virtual reality (VR) technologies offer particularly immersive platforms for emotional engagement. Our analysis identified four studies that harness VR's unique capabilities to create environments where users actively externalize, process, and transform emotional experiences \cite{10.1145/3613904.3642123,10.1145/3585088.3589381,10.1145/3460418.3479342,10.1145/3643834.3661570}.
These expressive making environments integrate other therapeutic approaches—exposure therapy \cite{10.1145/3460418.3479342,10.1145/3613904.3642123,10.1145/3585088.3589381}, mindfulness practice \cite{10.1145/3613904.3642123}, and cognitive reappraisal \cite{10.1145/3460418.3479342,10.1145/3643834.3661570,10.1145/3585088.3589381}—while enabling users to create, manipulate, and transform symbolic representations of their emotional states. This creative agency manifests across diverse contexts, from gamified therapeutic VR gardens symbolizing healing journeys \cite{10.1145/3613904.3642123} to home-based applications like VR Mood Worlds \cite{10.1145/3460418.3479342} and MoodShaper that encourage self-directed emotional expression and reflection \cite{10.1145/3643834.3661570}. Developmental applications for adolescents further demonstrate how symbolic creation and manipulation of emotional scenarios can facilitate cognitive reappraisal skill development \cite{10.1145/3585088.3589381}, highlighting expressive making's versatility as an intervention mechanism across different populations and contexts.

\subsubsection{Self-Acceptance and Compassion}
Self-acceptance and compassion enable individuals to acknowledge feelings and recognize emotional experiences through validation. Our analysis reveals how sensory-rich technologies effectively facilitate and foster these mechanisms by providing in-situ support that acknowledges emotional states while encouraging gentle self-regulation.
The cultivation of self-acceptance manifests particularly through technologies simulating affective touch—including artificial hugs and soothing wearables \cite{10.1145/3341163.3346933, 10.1145/3173574.3174069, 10.1145/3569484}—which create physiological experiences of reassurance and comfort that reinforce self-validation. Robot companions further extend this mechanism by embodying supportive presence, creating opportunities for social interactivity and affective engagement that normalize emotional experiences \cite{10.1145/3434073.3444669, 10.1145/3660244, 10.1145/2544104}. In care environments, embodied technologies such as interactive cushions with textile sensors and touch-sensitive plush toys with embedded speakers provide situated support that encourages self-acceptance across different populations, from children to elders \cite{10.1145/3490149.3501326, 10.1145/3491102.3502135, 10.1145/3637409}. These technologies translate physical and contextual signals into responsive sensory feedback that promotes experiential emotion processing centered on self-compassion.
\begin{figure}[h]
        \centering
        \includegraphics[width=0.75\linewidth]{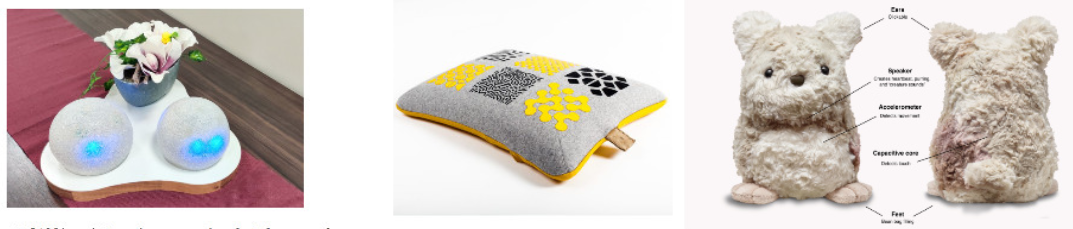}
        \caption{Examples of Self-Acceptance and Compassion emotion intervention approach that design tangible interfaces. From left to right: a tangible conversational agent designed for dementia patients \cite{10.1145/3490149.3501326}; an interactive cushion integrating textile sensors with touch-activated audio playback \cite{10.1145/3173574.3174069}; and a smart toy incorporating speakers and accelerometer technology \cite{10.1145/3491102.3502130}.}
        \label{fig:toyexample}
    \end{figure}
    
\input{tabs/annotation-MR}
\subsection{Goals of Technology-Mediated Emotion Interventions}
We analyzed the goals of empirical studies from two aspects in relation to their goals of emotion regulation and care-centered goals. 
\subsubsection{Emotion Intervention Goals: An Intervention Dichotomy} 
We identified an intervention dichotomy that emerged across 20 reviewed studies that aim to mitigate negative emotions and/or boost positive feelings. Such latent notions about what needs to be modified indicate a tendency to pathologize certain emotions, particularly those that deviate from societal norms.

\paragraph{Down-regulate negative and intense emotions}
12 studies were motivated by down-regulating negative, intense emotions in the moment, such as anxiety, distress, and anger, as well as controlling "inappropriate" displays of these emotions \cite{10.1145/3329189.3329209,10.1145/2809915.2809917}. Primarily supported by chat-based tools and sensory affordances, these endeavors drew on cognitive restructuring \cite{10.1145/3643834.3661570,10.1145/3613904.3642761}, cognitive reappraisal \cite{10.1145/3585088.3589381, 10.1145/3479561},  dialectical behavioral therapy \cite{10.1145/3421937.3421975} to help users reframe negative thoughts \cite{10.1145/3479561} or distance themselves from "non-adaptive" emotions \cite{10.1145/3643834.3661570}; While all 12 studies reported promising results on shifting emotional valence or decreasing intensity of negative affect over the course of deployment, individual experiences were more complex than reduction in metrics. For instance, the slow stroking by a anxiety-reduction wearable was deemed to signal consolation in competitive settings, which reinforces self-perception of weakness.  \cite{10.1145/3569484}. Other unintended outcomes of down-regulation include cognitive overload for consistently engaging in reappraisals \cite{10.1145/3479561}, unwanted focus on negative triggers during reframing \cite{oleary_suddenly_2018, 10.1145/3479561},  feeling uneasy and helpless due to tangible portrayal of negative emotions in immersive environments  \cite{10.1145/3643834.3661570} . In addition, two studies showed limited efficacy to transfer benefits of down-regulation into sustainable emotional support. For instance,  although peers' reframes tampered with negative thinking, few participants in \cite{10.1145/3479561} learned or independently applied these skills. Relatedly, visual manipulations of negative affect in VR helped users modulate challenging states but did not enhance clarity and awareness of their emotional experiences \cite{10.1145/3643834.3661570}.

\paragraph{Up-regulate positive affect and motivation}
Seven studies aimed to up-regulate positive feelings, such as pleasantness and social energy in everyday care environments and motivation to interact with intervention devices or systems \cite{10.1145/3575879.3575997, 10.1145/3173574.3174069, 10.1145/2674396.2674456, 10.1145/3152771.3152793, 10.1145/3154862.3154929, 10.1145/3490149.3501326, 10.1145/3613904.3642123, 10.1145/3660244}. This goal often emerged from design endeavors for populations with emotional expression barriers (e.g., the elderly with dementia, children with Autism Spectrum Disorder) \cite{10.1145/3154862.3154929, 10.1145/3490149.3501326, 10.1145/3575879.3575997, 10.1145/3152771.3152793}. Studies mostly operationalized this goal through game elements in designated  immersive environments (e.g., VR, interactive aquarium)   \cite{10.1145/3613904.3642123, 10.1145/3154862.3154929, 10.1145/2674396.2674456} or playful interactions sprinkled across everyday scenarios to record and explore emotions \cite{10.1145/3490149.3501326, 10.1145/3575879.3575997, 10.1145/3173574.3174069, 10.1145/3152771.3152793}. While researchers acknowledged the challenge of engaging these individuals in effective ER, who may have limited capacity to mobilize themselves at times of uncertainty or social encounters, the emphasis on a lively social presence and its assumed link with extraversion connote biased assumptions about the desirability of certain emotions in Western, agentic societies. Few studies gauged whether such up-regulation was desirable from the perspectives of end-users and their situational demands. Only three interventions were co-designed with care receivers to integrate their unique preferences and expectations\cite{10.1145/3490149.3501326, 10.1145/3152771.3152793, 10.1145/3660244}; two other studies sought feedback from caregivers (e.g., therapists, nursing home staff) as "experts" to inform user needs without hearing directly from whom would receive care \cite{10.1145/3613904.3642123, 10.1145/3154862.3154929}.

\subsubsection{Care-Centered Goals}

Two care-centered goals of the empirical studies emerged, non-judgmental emotion support and identity preservation. 

\paragraph{Non-judgmental emotion support}

Few studies proactively embraced the broad spectra of emotions in setting up intervention goals, which entail not only acknowledging positive and negative states but also leaving room for ambivalent, dynamic emotional experiences. For instance, one intervention allowed children with ADHD to log multiple emotions via different emoji combinations to indicate how they were feeling on the tangible home displays \cite{10.1145/3628516.3655795}. Another study sought to "sensitively support" young female patients diagnosed with cancer to embrace the array of negative emotions as they contended with the decisions on fertility preservation (e.g., distress, anger, upset, regret) \cite{10.1145/3639701.3656304}. Relatedly, three studies took the initiative to create a safe space for users to release unpleasant emotions and share negative experiences \cite{10.1145/3585088.3589381,10.1145/3613904.3641922, 10.1145/3411764.3445178}. Such efforts to foster self-compassion and safe emotional sharing allowed users to feel their feelings instead of suppressing or reconstructing the states presumed to undermine emotional well-being. 


In addition, only two studies reflect temporal dynamics of emotion in their approaches to ER support. For instance, through interactive videos that offer validation via the lived experiences of other patients, \cite{10.1145/3639701.3656304} assured users that it is okay for emotions to change across stages of medical decision-making, as they impose different cognitive and emotional challenges. Further, \cite{10.1145/3109761.3109790} sought to model the evolution of patients' emotions as they go through treatment procedures to inform caregivers to respond appropriately, especially in painful and sensitive care treatments. 


\paragraph{Identity preservation}
Eight studies aimed to empower individuals to preserve their identities amid health challenges \cite{10.1145/3491102.3502135, theofanopoulou_exploring_2022, 10.1145/3411764.3445178, 10.1145/3329189.3329209, 10.1145/3415183, 10.1145/3628516.3655795, 10.1145/3154862.3154929, 10.1145/3152771.3152793}. For instance, interventions for the elderly with dementia set out to kindle "reacquaintance with the self", helping them reenact their identities in the race against cognitive decline and emotional inertia \cite{10.1145/3490149.3501326, 10.1145/2544104}. Designs intended for youth with ASD and ADHD aimed to help individuals develop and stabilize a sense of personhood, by equipping them with fluid conduits to understand and express emotions \cite{10.1145/3411764.3445178, 10.1145/3329189.3329209, 10.1145/3152771.3152793}.

In addition, five studies sought to embed designs within care environments from inception to enrich the everyday experiences of individuals. As daily rhythms are integral to identity construction during the care journey, researchers aimed to mitigate the risk of ER interventions disrupting people's routines, or offered strategies to mend existing fractures and address related emotional challenges. For instance, \cite{10.1145/3544548.3581048} developed a gamified mobile app for daily planning to help individuals with autism cope with their anxiety that was intensified by broken routines. Through tangible objects with multisensory affordances, \cite{10.1145/3490149.3501326} provided elder residents with opportunities for open, playful engagement in their everyday moments across the care home space. Rather than assuming the contexts in which emotions were expressed, three studies also gathered contextual signals via sensors and social robots in care environments to paint a comprehensive emotional landscape for each user by accounting for their daily activities \cite{10.1145/3511047.3537691, 10.1145/3383313.3412244, 10.1145/2674396.2674453}.

\subsection{Roles of Technology-Mediated Emotion Interventions in Care Work}
Analysis of our corpus, viewed through critical examinations of care production, revealed three distinct roles that technology-supported emotion intervention tools served in care contexts: generating artificial care, nudging behavior changes, and mediating power dynamics. These roles emerged from examining how different technological implementations addressed various aspects of emotional support needs in caregiving relationships. 

\subsubsection{As a generator of artificial care}
Our corpus analysis revealed extensive research into artificial care—defined as non-human-generated emotional support, including simulated hugs and algorithmic empathetic responses. These studies examined how such implementations could address the labor-intensive nature of providing continuous emotional support in caregiving contexts.

Virtual Reality (VR) emerged as a significant focus area, with multiple studies \cite{10.1145/3643834.3661570, 10.1145/3563657.3595998, 10.1145/3491102.3517492, 10.1145/3357236.3395560} investigating immersive environments for emotional support delivery. Researchers explored various tangible devices \cite{10.1145/3410530.3414375, 10.1145/3173574.3174069, 10.1145/3491102.3502135}, developing rich sensory interactions specifically designed to elicit emotional responses. Studies of chatbot implementations \cite{10.1145/3589959} documented the development of conversational interfaces programmed to provide readily available empathetic responses.

The corpus documented both the potential benefits and specific limitations of artificial care systems. While studies showed promise in reducing caregivers' emotional labor through constant availability, researchers identified several challenges. These included difficulties in adapting to nuanced emotional contexts, potential risks of users developing unrealistic attachments to artificial care systems, and recurring concerns about the authenticity of machine-generated empathy. 
In response to these challenges, several studies explored integrated approaches. Research on AI-assisted suicide detection and emotional support systems for online counselors \cite{10.1145/3613904.3641922} documented attempts to enhance human capabilities in providing responsive and empathetic care while maintaining human oversight of the care process.

\subsubsection{As a mediator to intervene in power dynamics}
Studies in our corpus extensively documented how technology is being implemented to address inherent power dynamics in caregiving contexts, particularly focusing on situations where traditional power imbalances lead to emotional suppression among vulnerable individuals. We identified a cluster of technological interventions designed to create more equitable emotional expression opportunities within care relationships.

Multiple studies recognized that emotional coping needs frequently originate from power imbalances in daily interactions. Research by \textcite{10.1145/3610073} specifically examined romantic relationships, documenting how technology was implemented to address contraception responsibility imbalances. Their findings detailed how technological interventions created new channels for couples to discuss birth control and emotional support needs, working to increase mutual accountability in these conversations.

In examining parent-child relationships, where children's emotional expression capabilities may be limited, researchers explored novel technological approaches. \textcite{theofanopoulou_exploring_2022} documented the implementation of embodied, in-situ toys, studying how these devices could serve as alternative channels for children to communicate their emotional experiences to parents. Their research examined how such technological interventions could potentially equalize emotional exchanges between parents and children.

Studies in professional care settings documented efforts to enhance care receivers' visibility and agency. Researchers investigated tools designed to improve caregivers' emotional recognition capabilities \cite{10.1145/3575879.3575997} and support systems \cite{10.1145/3415222}, examining how these implementations could indirectly empower care receivers by ensuring better recognition and addressing of their emotional needs.
Research in dementia care settings \cite{10.1145/3490149.3501326} documented specific challenges to medicalized approaches, examining how technology could support the reconceptualization of care homes as residential spaces rather than clinical environments. These studies investigated how technological implementations could simultaneously support resident autonomy while maintaining necessary care provision, recognizing the intrinsic connection between emotional well-being and individual agency.

\subsubsection{As a nudge for action/behavior change}
Our analysis revealed numerous studies examining technology's role in maintaining emotional awareness within caregiving contexts where physical demands often overshadow emotional needs. Researchers investigated various implementations of tangible devices and self-tracking applications \cite{10.1145/3491102.3501925, 10.1145/3429360.3468190, 10.1145/3152771.3152793, 10.1145/3643834.3661570, 10.1145/3628516.3655795}, documenting how these tools helped users identify and label their affective states in relation to daily experiences. Studies examined specific implementations for different user groups, including emotion visualization tools for autistic individuals to aid in interpreting emotional cues \cite{zolyomi_emotion_2024}, and caregiver-tracking systems designed to inform clinical recommendations for managing challenging behaviors in children \cite{10.1145/3512939}.

The corpus documented attempts to intervene earlier in the emotion regulation (ER) process, examining approaches that moved beyond immediate emotional state modification to foster long-term awareness and skills. While empirical evaluation of these approaches remained limited, several studies employed established measurement tools. Researchers utilized the Difficulties in Emotion Regulation Scale \cite{10.1145/3643834.3661570}, measured changes in emotional controllability beliefs \cite{10.1145/3491102.3502130}, and evaluated the perceived learnability of ER skills post-intervention \cite{10.1145/3613904.3642761}. Qualitative analyses of user evaluations documented increased reflection and awareness of emotions and negative thought patterns \cite{10.1145/3628516.3655795, 10.1145/3479561, 10.1145/3491102.3501925}, suggesting potential improvements in emotional well-being and interpersonal functioning within caregiving relationships.

\section{Discussion}
In this paper, we presented a scoping review of prominent HCI publication venues in the ACM Digital Library. We identified 53 empirical studies that deploy technology-mediated emotion interventions (TMEIs) in care contexts. Informed by critical theories, we examined the approaches, goals, and roles of these technologies. Our findings highlight the importance of "emotion support" that calls for human-centered emotion intervention as an overarching value, offering an alternative to the "emotion regulation" paradigm. In the following sections, we first conceptually discuss what "emotion support" means and how it realigns with the critical scholarships; based on this, we propose "emotion support" principles and design implications for TMEI in facilitating caregivers, care-receivers, and relationship-building in the care contexts in 5.2.

\subsection{Towards Human-Centered Emotion Support}
\one{
We fully acknowledge the efficacy of emotion regulation theories and approaches in contexts where individuals have clear expectations about their ideal emotional states \cite{gross2015emotion, scarantino2024emotion}. Pioneers of the regulation model, such as psychologists \textcite{gross2003individual}, have also emphasized the importance of emotion expression—aspects that align with our notion of emotion support but have received less attention from HCI scholars. Building on this foundation, we propose and foreground the "emotion support" approach for situations where individuals cannot select their circumstances—such as when coping with chronic, rare, or incurable illnesses—and when they experience emotions that are vague, uncertain, and dynamic. Rather than emphasizing control or modification of emotional responses, emotion support creates space for the full spectrum of emotional experiences—however vague, contradictory, or evolving they may be—validating their legitimacy without imposing expectations for resolution or transformation. This approach recognizes that emotions in such contexts exist not as discrete states to be regulated toward an ideal but as fluid, ambiguous experiences that reflect the profound uncertainty of the human condition. We untangle two key meanings of emotion support as suggested by our findings. By embracing this perspective, we call for technologies for emotion support that are more nuanced, inclusive, and effective.}

\subsubsection{Center Humans' Diverse Emotion Support Needs}

In 4.1, we analyzed the goals of TMEIs and uncovered an intervention dichotomy that reflects the constant push for feeling joyful, sociable, and energetic in extant technology designs, which aimed to "up-regulate" individuals with interactive, sensory-rich stimuli. This approach may neglect people's diverse emotional needs and varied capacities to engage in interventions, especially in care contexts. Psychologists suggest that inertia does not always indicate "unwell" \cite{rzeszutek2021daily, nesse2019good}. Populations prone to emotional volatility due to health complications might benefit from the private space to process feelings or enjoy solitude, which allows them to attend to and explore the array of emotions as they naturally arise. Relatedly, the predominant down-regulation goal in our corpus may inadvertently silence the generative roles of negative emotions. However, psychology literature has long recognized the adaptive functions of "negative" affect, for instance, anxiety and anger may sharpen focus and fuel motivation \cite{tamir_why_2016, gross_emotion_2015, nesse2019good}. Translating these insights into HCI design, more interventions are needed to help users understand, validate, and learn from their emotional experiences. 

Beyond the intervention dichotomy, we observed two care-centered goals in our corpus: non-judgmental support for all shades of emotions and identity preservation across the care journey. We highlight studies that keenly attend to the interplay between the "person" and "context" - individuals have unique, dynamic emotional needs in their care setting \cite{10.1145/2544104, 10.1145/3329189.3329209, 10.1145/3152771.3152793}; it is crucial to identify when and how technologies could intervene to deliver sensitive emotion support, without compromising individuals' evolving identities and their rich emotional life. Critical scholars also view emotions as a source of knowledge and power rather than weakness \cite{boehner2007emotion, howell2018emotional, lin2021techniques, ahmed2013cultural}. This perspective encourages validating the full spectrum of emotions, recognizing their importance in decision-making and social change. Such validation-driven approaches may foster personal growth and de-escalate intense affect when individuals find it necessary, which allows them to engage with affective signals on their own terms as they navigate care decisions \cite{kashdan2010psychological}. To this end, our work echoes the commitment outlined by critical scholars in HCI to shift the goals of sociotechnical systems from "managing" to "nurturing" people \cite{bardzell_towards_2011, lin2021techniques}; the journey of negative emotions can be uplifting if guided with sensitive, identity-preserving technology designs.

\paragraph{Methodological Implications:}
%

\one{
We advocate for co-constructing the goals and approaches to designing technology-mediated emotion support, which empowers individuals to self-define how they want to experience their emotions~\cite{bardzell_towards_2011}. This calls for a methodological turn for researchers to truly appreciate and respect individuals' emotional needs.
Notably, Bennett and Rosner \cite{bennett2019promise} surfaced the pitfalls of researchers' "empathy activities" in displacing disabled users from their lived experiences, instead proposing the ethos of "being with", not "being like" the intended users throughout the design process. This requires researchers to deliberately recognize and work with the asymmetries between intended users and the intervention device, which fosters ongoing attunement with their dynamic needs for support.
}

\one{
Hence, we call for critical reflection on whose stories to account for and how to authentically represent them when developing interventions. Researchers should actively learn about and stay informed of how intersecting identities of individuals shape their emotional experiences and their distinct needs for emotion support \cite{10.1145/3078072.3079742, 10.1145/3359187}. We join Bennet and Rosner \cite{bennett2019promise} and Williams et al. \cite{williams2023counterventions} to foreground participants as partners in envisioning design encounters, and cultural informants in contextualizing the design "problem" within their social realities. While co-design workshops may uncover insights on what feeling "better" entails for different user groups, and online communities offer vast data sources to learn from authentic narratives of diverse populations, we also highlight offline partnerships as ongoing commitment for researchers and designers to embed oneself within the local practices of marginalized individuals, whose voices tend to be drowned out by power differentials \cite{pendse_treatment_2022}. For example, by collaborating with women who have lived through historically understudied and often stigmatized health conditions, scholars have expanded research into previously overlooked areas such as postpartum depression and endometriosis \cite{kumar_womens_2019} and menstrual health technologies \cite{fox_imagining_2017}.
}

\one{
Drawing on critical scholarships in HCI that challenge the normative assumptions of well-being \cite{adler2022burnout, docherty2022re, jardine2024between}, we highlight the need for researchers to resist curative framings of interventions \cite{williams2023counterventions, stramondo2019distinction} and embrace the necessity of negative emotions in care receivers' recovery journey \cite{kashdan2010psychological, md_good_2019,gross_emotion_2015}. Beyond sociocultural norms of what and how emotions "should" be expressed, participatory designs may strive to elevate the agency of individuals to explore and process intense, negative feelings, for instance, through playful experiments with accessible, familiar materials in their everyday environments to easily craft sensory narratives of negative emotions. This approach preserves the breathing space of individuals at challenging moments, while mitigating potential harms of unintended negativity focus through prolonged engagement.
}

\subsubsection{Honor Experiential and Relational Paradigm}
\one{
To truly acknowledge and support the diverse emotional needs of individuals, emotion support approach calls for experiential and relational paradigms that transcend cognitive and individual mechanisms as reflected in the current TMEI practices. 
}

\one{
Our analysis of the current emotion technology landscape reveals an imbalance: the reviewed studies predominantly feature interventions targeting cognitive or behavioral manifestations of emotions (i.e., cognitive change and response modulation in Section 4.1), while designs that engage with the experiential dimension of emotional processes remain rare. This gap becomes particularly notable when considering perspectives like \textcite{damasio1994descartes}, who challenges the Cartesian separation of mind and body by demonstrating that emotions are essential components of rational thinking and cognitive development—not secondary phenomena to be controlled or eliminated. Such perspectives suggest fundamental limitations in approaches focused primarily on cognitive modulation of emotions.
While many studies are rooted in Gross's process model of emotion regulation \cite{gross_emerging_1998, gross_emotion_2007, gross_handbook_2015}, we also identified several promising approaches that embody the concept of "emotion support," such as social support, expressive making, and acceptance as listed in Table \ref{tab:EImechanism}.
These emotion support approaches are grounded in diverse and well-established psychological traditions that offer rich alternative frameworks. These include Rogers' humanistic psychology \cite{rogers1951client, rogers1995becoming}, which emphasizes unconditional positive regard; attachment theory developed by Bowlby \cite{bowlby2008secure}, which highlights the importance of secure emotional bonds; and Linehan's dialectical behavior therapy \cite{linehan1993cognitive}, which integrates validation with change strategies. These psychological frameworks collectively argue that emotional validation—rather than suppression or purely cognitive reframing—promotes secure attachment and constitutes an essential component of effective therapy. The efficacy of these approaches has been subsequently confirmed by empirical studies \cite{rauschenberg_compassion-focused_2021, ford2018psychological}. These findings suggest an opportunity to expand and diversify the theoretical psychological foundations that inform HCI research and practice.
}

\one{
Embracing design justice principles \cite{costanza-chock_design_2020}, the "emotion support" paradigm addresses collective liberation by acknowledging sociocultural contexts that shape emotional processing across diverse communities. Design justice advocates for community-led practices centering on the experiences of those "multiply burdened under the matrix of domination." However, our review reveals that most technology-mediated emotion interventions (TMEIs) operate at the individual level, leaving relational approaches underdeveloped. This individualistic focus risks modulating emotional experiences in isolation, undermining solidarity, especially given the pervasiveness of technology-delivered interventions.
Cultural models of emotion \cite{mesquita2007cultural} emphasize that many societies manage emotions socially rather than through individual suppression or reappraisal, calling for a paradigm shift from private regulation to relational and communal processing. Design justice principles suggest reconceptualizing interventions by integrating community-defined well-being practices, particularly for marginalized groups where collective coping has historically been essential for survival. For example, Kim et al. \cite{kim2024envisioning, kim_design_2025} introduced community-based positive psychology approaches that offer promising alternatives that emphasize collective strengths flourishing \cite{to_flourishing_2023} over individualistic happiness concepts.
}

\paragraph{Methodological Implications:}
\one{
We advocate for future designers to embrace a more diverse spectrum of psychological theories that recognize the embodied nature of emotions—theories that acknowledge the complex interconnections between physical sensations, cognitive processes, and emotional experiences \cite{elizabeth1994volatile}. Researchers in this space should approach psychological theories with greater reflexivity, drawing on critical scholarship to innovate and recenter human experience in their theoretical applications.
Undoubtedly, established emotion regulation frameworks such as Gross' model have yielded valuable outcomes in HCI research. Slovak et al.'s \cite{petr_slovak_designing_2023} agenda-setting work exemplifies this contribution by systematically mapping emotion regulation implications across theoretical, strategic, and practical dimensions, thereby providing psychology-informed foundations for emotion regulation research in HCI. 
However, we advocate for expanding beyond these established models to create opportunities for more nuanced and comprehensive approaches to human-centered design.
Adding a reflexive lens can deepen research and expand toolkits for understanding the "situated knowledge" of people as cultural informants and knowledge sources. By critically engaging with contemporary trauma research, trauma-informed design \cite{chen_trauma-informed_2022, scott_trauma-informed_2023} identifies emotional support—characterized by compassionate behaviors offering encouragement and reassurance—as a crucial positive social reaction facilitating recovery for trauma survivors.
Furthermore, by engaging with diverse cultural understandings of mental wellbeing beyond historically Western psychological frameworks, HCI researchers can explore more holistic approaches to well-being \cite{smith_designing_2024}, including how spiritual practices \cite{smith_what_2021} may support resilience across different communities.
}

\one{
To truly design interventions that foster experiential and relational emotion support, we advocate for technologies that afford users the agency and rich pathways to construct their emotions as interleaved with thoughts and senses and to reflect on how sociocultural norms shape such malleable experiences.
Designing experiential systems for emotion support requires embracing the plurality of human bodies and emotional experiences. Research suggests that effective designs balance ambiguity and clarity to facilitate embodied interaction with one's own feelings and those of others \cite{costanza-chock_design_2020, 10.1145/1094562.1094570}. This approach centers users' flexible interpretation of their bodily signals while offering comfort and agency when technologies access intimate personal spaces \cite{grill_attitudes_2022}. 
Building upon this foundation, researchers can move beyond viewing the data in emotion tracking technologies as objects to embrace various ways of caring-through-data versus data-as-care \cite{kaziunas_caring_2017}, such as biodata to foster shared living and knowing, creating alternative forms of care \cite{10.1145/3532106.3533477, 10.1145/3130943}. 
For instance, \textcite{10.1145/2901790.2901850} designed ambiguous clothing-based displays of skin conductance that validate personal feelings while providing social cues to others, enabling collaborative interpretation and meaning-making around emotional signals. These designs transcend simplistic categorical emotion labeling, instead creating experiential systems that honor the subjective, fluid nature of emotional experiences while fostering interpersonal connection and mutual understanding.
}

\subsection{Situating roles of technologies within the socio-tech ecosystem}
\one{
We identified various roles that technologies serve in technology-mediated emotion interventions in the carework contexts. This section outlines implications for future design in developing technologies to support various actors in the sociotechnical care ecosystem. 
}
\subsubsection{Design for Caregivers: Foster Human Care and Connections}
Our review reveals a problematic trend in current TMEIs: the deployment of sophisticated computational methods to create algorithm-driven "artificial care" as an alternative—and in many cases, a replacement—for human care. Many studies attempt to use algorithms to label people's care needs or generate empathetic responses in conversations \cite{10.1145/3613904.3642761, 10.1145/3442381.3450097}, particularly when human caregivers are deemed unavailable or their involvement is considered burdensome.
While machine-generated artificial empathy can alleviate health caregivers' emotional labor and potentially reduce burnout \cite{smriti_bringing_2023}, a delicate balance must be maintained regarding which aspects of care should remain human-centered. Evidence suggests that caregivers prefer to preserve meaningful emotional interactions with care-receivers while hoping technologies might handle more mundane and transactional affairs such as completing paperwork and communicating with insurance companies \cite{smriti_bringing_2023, smriti_emotion_2024}.
Paradoxically, while the logistical aspects of care work remain underaddressed in research \cite{llubes-arria_emotional_2022}, more studies attempt to replace precisely those emotion-intense interactions that caregivers themselves value most \cite{schorch_designing_2016, smriti_emotion_2024}.

From an emotional support perspective, we argue that human care remains irreplaceable, and the shift toward "artificial care" mischaracterizes care's fundamental nature. This technological trajectory overemphasizes technologies' supposed capacity to bypass structural and social barriers, reducing complex but authentic relational connections to algorithmic interactions detached from their social contexts. Voluntary human caring should not be colonized by market forces and commercial interests that commodify individualistic self-care applications \cite{spors_selling_2021}.
Care provision—encompassing organizational, physical, and emotional labor \citep{hochschild2019managed, smriti_emotion_2024}—is intrinsically relational and cannot be effectively replicated through artificial means. In the cultural analysis of emotion, \textcite{ahmed2013cultural} emphasizes that authentic emotional support requires shared spaces, dedicated time, and genuine human interaction. Machine-generated "empathy," devoid of lived experience, risks not only ineffectiveness but potentially deterring meaningful human communication. Rather than designing technologies that attempt to supplant human caregivers, we should leverage technology to enhance and sustain the irreplaceable value of human connections in care contexts.


\paragraph{Design Implications:}
Following the two principles of designing human-centered interventions that honor emotion support, we propose reframing empathy not as a message, but as a deliberate commitment—one that can be facilitated through technology-supported attunement in care work. This perspective preserves the irreplaceable value of human connection while addressing the challenges of sustainable caregiving. While algorithm-generated empathy and cognitive reappraisals have their place, embedding them within a human-moderated framework is essential. For example, \textcite{10.1145/3613904.3641922} used LLM-assisted message generation to support moderators on online forums to communicate with people experiencing suicidal ideation. This approach not only preserves the human-to-human connection vital to emotional support but also addresses pragmatic concerns, ensuring that unpredictable situations and emerging malfunctions—particularly common in care contexts where extreme and sensitive emotions frequently arise—can be handled appropriately. Relatedly, we highlight the potential of using linguistic intelligence to connect people, facilitating support through algorithm-mediated communication such as referring to others' stories and experiences, providing direct and navigational help \cite{fang_matching_2022}.


\subsubsection{Design for Care-Receiver: Preserve Identity Across Care-Seeking Journey}

Our review highlights a growing body of research on emotion support that addresses the intersected marginality of individuals with diseases,  which complicates their emotional experiences throughout the healing journey. Rather than aiming to "regulate" emotions, these studies emphasize preserving the identities of care-receivers and validating the full spectrum of emotions that arise with their identity shifts. Identity is deeply intertwined with how people experience and cope with illness, influencing their symptom recognition, disease interpretation, self-reflection, and help-seeking behaviors \cite{pelttari_emotional_2022}. In the face of life-altering diagnosis and uncertain prognosis, patients often grapple with anxiety, fear, and depression, compounded by loss of control when forced to make sense of their new realities.  As evident in our corpus, broken routines intensified the anxiety of autistic individuals; sensory barriers to engage with the world destabilized elders' sense of self \cite{10.1145/3544548.3581048, 10.1145/3490149.3501326}; young female cancer patients struggle to reconcile the societal expectations of womanhood with their treatment decisions related to fertility preservation \cite{10.1145/3639701.3656304}.


Recognizing such dynamic identity perceptions in healthcare contexts, our review underscores personalized, identity-sensitive support in designing care technologies. Superficial displays of digital empathy, often provided by chatbots and LLMs, may overlook the unique personal contexts and social realities of care-receivers, risking emotional harm rather than offering authentic support that acknowledges their intersecting identities \cite{10.1145/3078072.3079742, ahmed2013cultural, cuadra2024illusion, roshanaei2024talk}.
Further, traditional emotion regulation strategies such as situation selection and modification may be infeasible or inadequate in cases of severe or life-limiting illnesses, or even counterproductive when one has little control over their environment and care trajectories \cite{smith_digital_2022, goodley_feeling_2018}. Thus, we foreground interventions embedded in daily routines and disease journeys in 4.2.2 \cite{10.1145/3544548.3581048, 10.1145/3490149.3501326, 10.1145/3511047.3537691, 10.1145/3383313.3412244, 10.1145/2674396.2674453} - technologies should adapt to varied energy, motivation, and goals of care-receivers for engaging with their emotions, as they settle into new identities with symptom development \cite{zhang_designing_2021}. We also highlight the studies in 4.3.2 that explored alternative channels of expression to empower care-receivers, especially in relationships inscribed with unevenly distributed responsibilities (e.g., romantic partners, children and parents, care home residents and caregivers) \cite{10.1145/3610073, theofanopoulou_exploring_2022, 10.1145/3575879.3575997}. Future work can build on these approaches by embracing diverse forms of emotional exchanges, especially for individuals whose health conditions and psychological states limit their verbal communication.

\paragraph{Design Implications:}
Designers should create space for individuals to document and reflect on their pre-illness identities and evolving self-perceptions. Narrative-based features such as digital journals and interactive timelines can promote meaning making and identity reconstruction along symptom development; gradual goal-setting frameworks can also align interventions with the shifting capabilities of care receivers. These strategies may help individuals navigate the emotional challenges rooted in their lived experiences, sparking sustainable behavior changes that resonate with their emerging identities\cite{10.1145/3544548.3581048, 10.1145/3628516.3655795}. 




Joining HCI scholarships on critical making \cite{ratto2011critical, grimme2014we, iivari2018empowering} and the "making as expression" perspective inspired by art therapy \cite{10.1145/3359187}, we call for non-verbal means to empower populations with communication barriers in collaborative design, such as individuals with aphasia, dementia, and traumatic experiences. Notably, Lazar et al. \cite{10.1145/3359187} suggest attending to unique languages of materials (e.g., fabrics, clay, glass) to help individuals articulate their thoughts and feelings. Similarly, Hong et al. \cite{hong2018visual} found visualizing emotions via storyboards to be a comfortable alternative for teens with chronic illness to communicate with their parents. Researchers should draw on materials beyond verbal modalities to facilitate care-receivers in fluid, embodied expressions of their identities and emotional needs.

\subsubsection{Design for Care Relationships: Foreground Power Dynamics in Carework}
Our review highlights the need for TMEIs to directly address power dynamics among caregiving stakeholders. From an emotion support perspective, we illuminate three critical tensions sidelined by dominant caregiving paradigms: (1) temporal disconnects between the rigidity of institutional care schedules and the fluid, unpredictable nature of emotional needs—as evinced in 4.2.2, few interventions accounted for dynamics of emotions as individuals navigate medical decisions and treatment procedures; (2) unidirectional conceptualizations of care that reinforce colonial notions of "delivery", rather than recognizing indigenous and feminist traditions of care as co-constructed \cite{boiger2012construction}—this tension manifests in relatively uncharted roles of care technologies in mediating power imbalance, and lack of support for caregivers' emotional needs in 4.3; (3) siloed emotional experiences that reflect neoliberal individualism rather than collective approaches to emotional wellbeing—most interventions emphasize self-management through emotion tracking (4.1), with technologies mainly serving as a personal nudge (4.3).

In response, we call for future designs to embrace a radical re-conceptualization of emotion support in care contexts. Researchers must acknowledge how structural factors—including racism, sexism, classism, and ableism—shape the visibility of emotional labor on part of all care stakeholders \cite{gorton2007theorizing, jaggar2014love}. This approach centers the fundamental power imbalances within care ecosystems, ensuring that both caregivers and care receivers from marginalized communities have their expertise recognized, their agency honored, and their emotional labor valued.
By centering the knowledge of those historically excluded from care design, transformative socio-technical systems can move beyond efficiency to actively disrupt hierarchical exchanges. Rather than reinforcing techno-solutionist approaches, designs informed by critical theory should embrace caregiving's complex, relational nature to create more equitable care relationships \cite{crenshaw2013mapping}. 


\paragraph{Design Implications:}

Technologies addressing these power imbalances must foster mutual emotional engagement rather than merely expediting care tasks. By embedding caregivers earlier in emotional journeys \cite{10.1145/3490149.3501326} and prioritizing collaborative interpretation over automation \cite{10.1145/3152771.3152793, 10.1145/3512939,10.1145/3628516.3655795}, designs can redistribute emotional visibility and challenge knowledge hierarchies. This approach validates both care receivers' needs and caregivers' emotional engagement while promoting shared reflection through lived experiences, ultimately recentering human connection in care relationships.


Systems should connect users with similar care experiences to foster genuine community building beyond clinical contexts. Such technology-mediated connections should incorporate intersecting identities beyond diagnosis alone, recognizing that emotions are negotiated within complex social contexts. This approach validates the temporal dynamics of identity shifts throughout care journeys and facilitates collective coping \cite{10.1145/1517744.1517804}, creating more nuanced care environments that resist medical reductionism. Recent socio-technical interventions demonstrate this evolution through collaborative management tools \cite{burgess_i_2019}, group sense-making platforms \cite{andalibi_sensemaking_2021}, and stigma reduction initiatives \cite{siddiqui_exploring_2023}. These interventions intentionally balance power dynamics by creating spaces where multiple perspectives hold equal value and collective wisdom emerges organically. By distributing emotional labor across networks beyond dyadic relationships, these community-building tools can simultaneously reduce caregiver burden and expand agency for care receivers, directly addressing the power imbalances inherent in traditional care models.




\section{Conclusion}
Through a review of 53 empirical studies on Technology-Mediated Emotion Interventions (TMEIs), this work seeks to expand beyond traditional emotion regulation practices. We propose "emotion support" as an alternative approach that emphasizes non-judgmental care and identity preservation rather than simply regulating emotions. By examining these technologies within care ecosystems, we uncover their roles in providing artificial care, influencing power dynamics, and shaping behavior. These findings offer guidance for designing emotional support technologies that respect emotional complexity, maintain user agency, and enhance care relationships. We encourage HCI and CSCW researchers to consider more holistic approaches to emotional well-being that acknowledge the social nature of emotions within broader care contexts.

\bibliographystyle{ACM-Reference-Format}
\bibliography{extra, litreview-edit, mentalhealth-edit, acm321-edit}

\appendix

\end{document}

%% file: premeable.tex
\usepackage{csquotes}
\makeatletter
\@ifclasswith{acmart}{anonymous}
{%
    \usepackage[true]{anonymous-acm}
}{%
    \usepackage[false]{anonymous-acm} 
}
\makeatother
\usepackage{ifdraft}
\usepackage{etoolbox} 
\usepackage{xspace}
\usepackage{color}
\usepackage{tabularray}
\usepackage{comment}
\usepackage{cleveref}
\renewcommand*{\cref}{\Cref}
\usepackage{hyphenat}
\usepackage{array}
\usepackage{booktabs}
\usepackage{tabularx} 
\usepackage{multirow}
\usepackage{multicol}
\usepackage{utfsym}
\definecolor{lightyellow}{RGB}{255, 251, 204}

\usepackage{longtable}
\usepackage{colortbl}
\usepackage{multirow}
\usepackage{subcaption}
\usepackage{csquotes}

\NewDocumentCommand{\pquote}{+O{} +m}{\blockquote[#1]{\textit{#2}}}

\newcommand{\elide}[1]{\textelp{}} 
 
\newcommand{\textcite}[1]{\citeauthor{#1}~\cite{#1}}

%% file: tabs/EImechanisms.tex
\begin{table}[h]
\small
\caption{Emotion Intervention Approaches}
\label{tab:EImechanism}
\begin{tabular}{@{}p{2cm}p{6cm}p{6cm}@{}}
\toprule
\textbf{Approach} & \textbf{Definition} & \textbf{Related Theoretical Frameworks} \\ \midrule

Emotion tracking and feedback & 
Prompt recognition and comprehension via digital records and sensory feedback & 
Process Model of Emotion Regulation \cite{gross_emerging_1998, gross_emotion_2007, gross_handbook_2015}, Feelings-as-information Theory \cite{schwarz2012feelings} \\

Cognitive model development & 
Educate users about regulation strategies and guide them towards cognitive change & 
Process Model of Emotion Regulation \cite{gross_emerging_1998, gross_emotion_2007, gross_handbook_2015}, Distanced Self-talk \cite{kross2014self, kross2017self} \\

Expressive making & 
Encourage external expression via symbolic creation and manipulation of emotional scenarios & 
Expressive Writing \cite{pennebaker1997writing}, Expressive Therapies \cite{malchiodi2013expressive}, Humanistic Psychology \cite{rogers1951client, rogers1995becoming} \\

Social connection & 
Enhance interpersonal emotional exchanges and access to social support networks & 
Buffering Hypothesis of Social Support \cite{cohen1985stress}, Cultural Model of Emotion \cite{mesquita2007cultural} \\

Self-acceptance and compassion & 
Foster validation and processing of emotional experiences in-situ and offer reassurance & 
Attachment Theory \cite{bowlby2008secure}, Self-compassion \cite{neff2003self}, Dialectical Behavior Therapy \cite{lynch2006mechanisms} \\

\bottomrule
\end{tabular}
\end{table}

%% file: tabs/annotation-MR.tex
\begin{table}[h]
\caption{Categories of the empirical studies regarding Approaches, Goals, and Roles}
\label{tab:annotation}
\resizebox{\textwidth}{!}{%
\begin{tabular}{@{}lllllllllllll@{}}
\toprule
 &
  \textbf{Approaches} &
   &
   &
   &
   &
  \textbf{Goals} &
   &
   &
   &
  \textbf{Roles} &
   &
   \\ \midrule
\textbf{Paper} &
  \textbf{Feedback} &
  \textbf{Cognitive} &
  \textbf{Social} &
  \textbf{Express} &
  \textbf{Acceptance} &
  \textbf{Down} &
  \textbf{Up} &
  \textbf{Non-jud} &
  \textbf{Ident} &
  \textbf{Generator} &
  \textbf{Mediator} &
  \textbf{Nudge} \\
\textcite{10.1145/3479561} &
  \cellcolor[HTML]{FCE8B2}* &
  \cellcolor[HTML]{FCE8B2}* &
  \cellcolor[HTML]{FCE8B2}* &
   &
   &
  \cellcolor[HTML]{F4C7C3}* &
   &
   &
   &
   &
   &
   \\
\textcite{10.1145/3411764.3445178} &
  \cellcolor[HTML]{FCE8B2}* &
  \cellcolor[HTML]{FCE8B2}* &
  \cellcolor[HTML]{FCE8B2}* &
   &
   &
   &
   &
  \cellcolor[HTML]{F4C7C3}* &
  \cellcolor[HTML]{F4C7C3}* &
   &
  \cellcolor[HTML]{B7E1CD}* &
  \cellcolor[HTML]{B7E1CD}* \\
\textcite{10.1145/3628516.3655820} &
  \cellcolor[HTML]{FCE8B2}* &
  \cellcolor[HTML]{FCE8B2}* &
   &
   &
   &
   &
   &
   &
   &
  \cellcolor[HTML]{B7E1CD}* &
   &
  \cellcolor[HTML]{B7E1CD}* \\
\textcite{10.1145/3421937.3421975} &
  \cellcolor[HTML]{FCE8B2}* &
  \cellcolor[HTML]{FCE8B2}* &
   &
   &
   &
   &
   &
   &
   &
  \cellcolor[HTML]{B7E1CD}* &
   &
  \cellcolor[HTML]{B7E1CD}* \\
\textcite{10.1145/3613904.3641922} &
  \cellcolor[HTML]{FCE8B2}* &
   &
  \cellcolor[HTML]{FCE8B2}* &
   &
   &
   &
   &
  \cellcolor[HTML]{F4C7C3}* &
   &
  \cellcolor[HTML]{B7E1CD}* &
   &
   \\
\textcite{10.1145/3637409} &
  \cellcolor[HTML]{FCE8B2}* &
   &
   &
  \cellcolor[HTML]{FCE8B2}* &
   &
   &
   &
   &
   &
   &
  \cellcolor[HTML]{B7E1CD}* &
   \\
\textcite{10.1145/3349537.3351907} &
  \cellcolor[HTML]{FCE8B2}* &
   &
   &
  \cellcolor[HTML]{FCE8B2}* &
   &
   &
   &
   &
   &
  \cellcolor[HTML]{B7E1CD}* &
   &
  \cellcolor[HTML]{B7E1CD}* \\
\textcite{10.1145/3491102.3502135} &
  \cellcolor[HTML]{FCE8B2}* &
   &
   &
   &
  \cellcolor[HTML]{FCE8B2}* &
   &
   &
   &
  \cellcolor[HTML]{F4C7C3}* &
  \cellcolor[HTML]{B7E1CD}* &
  \cellcolor[HTML]{B7E1CD}* &
   \\
\textcite{10.1145/3442381.3450097} &
  \cellcolor[HTML]{FCE8B2}* &
   &
   &
   &
   &
   &
   &
   &
   &
   &
   &
   \\
\textcite{10.1145/3613904.3642761} &
  \cellcolor[HTML]{FCE8B2}* &
   &
   &
   &
   &
  \cellcolor[HTML]{F4C7C3}* &
   &
   &
   &
  \cellcolor[HTML]{B7E1CD}* &
   &
   \\
\textcite{10.1145/3396868.3400898} &
  \cellcolor[HTML]{FCE8B2}* &
   &
   &
   &
   &
   &
   &
   &
   &
  \cellcolor[HTML]{B7E1CD}* &
   &
   \\
\textcite{10.1145/3152771.3152793} &
  \cellcolor[HTML]{FCE8B2}* &
   &
   &
   &
   &
   &
  \cellcolor[HTML]{F4C7C3}* &
   &
  \cellcolor[HTML]{F4C7C3}* &
   &
  \cellcolor[HTML]{B7E1CD}* &
  \cellcolor[HTML]{B7E1CD}* \\
\textcite{10.1145/3628516.3655795} &
   &
  \cellcolor[HTML]{FCE8B2}* &
  \cellcolor[HTML]{FCE8B2}* &
   &
   &
   &
   &
  \cellcolor[HTML]{F4C7C3}* &
  \cellcolor[HTML]{F4C7C3}* &
   &
   &
  \cellcolor[HTML]{B7E1CD}* \\
\textcite{10.1145/3575879.3575997} &
   &
  \cellcolor[HTML]{FCE8B2}* &
  \cellcolor[HTML]{FCE8B2}* &
   &
   &
   &
  \cellcolor[HTML]{F4C7C3}* &
   &
   &
   &
  \cellcolor[HTML]{B7E1CD}* &
  \cellcolor[HTML]{B7E1CD}* \\
\textcite{10.1145/2809915.2809917} &
   &
  \cellcolor[HTML]{FCE8B2}* &
   &
   &
   &
  \cellcolor[HTML]{F4C7C3}* &
   &
   &
   &
   &
   &
   \\
\textcite{10.1145/3154862.3154929} &
   &
  \cellcolor[HTML]{FCE8B2}* &
   &
   &
   &
   &
  \cellcolor[HTML]{F4C7C3}* &
   &
  \cellcolor[HTML]{F4C7C3}* &
   &
   &
  \cellcolor[HTML]{B7E1CD}* \\
\textcite{10.1145/3511047.3537691} &
   &
  \cellcolor[HTML]{FCE8B2}* &
   &
   &
   &
   &
   &
   &
  \cellcolor[HTML]{F4C7C3}* &
   &
   &
  \cellcolor[HTML]{B7E1CD}* \\
\textcite{10.1145/3660244} &
   &
  \cellcolor[HTML]{FCE8B2}* &
   &
   &
  \cellcolor[HTML]{FCE8B2}* &
   &
   &
   &
   &
   &
   &
  \cellcolor[HTML]{B7E1CD}* \\
\textcite{10.1145/3109761.3109790} &
   &
  \cellcolor[HTML]{FCE8B2}* &
   &
   &
   &
   &
   &
  \cellcolor[HTML]{F4C7C3}* &
   &
   &
   &
   \\
\textcite{10.1145/3491102.3501925} &
   &
  \cellcolor[HTML]{FCE8B2}* &
   &
   &
   &
   &
   &
   &
   &
   &
   &
  \cellcolor[HTML]{B7E1CD}* \\
\textcite{10.1145/3383313.3412244} &
   &
  \cellcolor[HTML]{FCE8B2}* &
   &
   &
   &
   &
   &
   &
  \cellcolor[HTML]{F4C7C3}* &
   &
   &
  \cellcolor[HTML]{B7E1CD}* \\
\textcite{10.1145/3544548.3581048} &
   &
  \cellcolor[HTML]{FCE8B2}* &
   &
   &
   &
   &
   &
   &
  \cellcolor[HTML]{F4C7C3}* &
   &
   &
  \cellcolor[HTML]{B7E1CD}* \\
\textcite{10.1145/3512939} &
   &
  \cellcolor[HTML]{FCE8B2}* &
   &
   &
   &
   &
   &
   &
   &
   &
  \cellcolor[HTML]{B7E1CD}* &
   \\
\textcite{10.1145/2674396.2674456} &
   &
  \cellcolor[HTML]{FCE8B2}* &
   &
   &
   &
   &
  \cellcolor[HTML]{F4C7C3}* &
   &
   &
   &
   &
  \cellcolor[HTML]{B7E1CD}* \\
\textcite{10.1145/2674396.2674453} &
   &
  \cellcolor[HTML]{FCE8B2}* &
   &
   &
   &
   &
   &
   &
  \cellcolor[HTML]{F4C7C3}* &
   &
   &
  \cellcolor[HTML]{B7E1CD}* \\
\textcite{10.1145/3173574.3173905} &
   &
  \cellcolor[HTML]{FCE8B2}* &
   &
   &
   &
  \cellcolor[HTML]{F4C7C3}* &
   &
   &
   &
  \cellcolor[HTML]{B7E1CD}* &
   &
   \\
\textcite{10.1145/3429360.3468190} &
   &
  \cellcolor[HTML]{FCE8B2}* &
   &
   &
   &
   &
   &
   &
   &
   &
   &
  \cellcolor[HTML]{B7E1CD}* \\
\textcite{10.1145/2851613.2851678} &
   &
  \cellcolor[HTML]{FCE8B2}* &
   &
   &
   &
   &
   &
   &
   &
   &
   &
  \cellcolor[HTML]{B7E1CD}* \\
\textcite{10.1145/3490149.3501326} &
   &
   &
  \cellcolor[HTML]{FCE8B2}* &
  \cellcolor[HTML]{FCE8B2}* &
   &
   &
  \cellcolor[HTML]{F4C7C3}* &
   &
  \cellcolor[HTML]{F4C7C3}* &
   &
  \cellcolor[HTML]{B7E1CD}* &
  \cellcolor[HTML]{B7E1CD}* \\
\textcite{10.1145/3130943} &
   &
   &
  \cellcolor[HTML]{FCE8B2}* &
  \cellcolor[HTML]{FCE8B2}* &
   &
   &
   &
   &
   &
   &
  \cellcolor[HTML]{B7E1CD}* &
  \cellcolor[HTML]{B7E1CD}* \\
\textcite{10.1145/3078072.3079742} &
   &
   &
  \cellcolor[HTML]{FCE8B2}* &
  \cellcolor[HTML]{FCE8B2}* &
   &
  \cellcolor[HTML]{F4C7C3}* &
   &
   &
   &
   &
  \cellcolor[HTML]{B7E1CD}* &
   \\
\textcite{10.1145/3639701.3656304} &
   &
   &
  \cellcolor[HTML]{FCE8B2}* &
   &
   &
   &
   &
  \cellcolor[HTML]{F4C7C3}* &
   &
   &
   &
   \\
\textcite{10.1145/3415222} &
   &
   &
  \cellcolor[HTML]{FCE8B2}* &
   &
   &
   &
   &
   &
   &
   &
   &
   \\
\textcite{10.1145/3415183} &
   &
   &
  \cellcolor[HTML]{FCE8B2}* &
   &
   &
   &
   &
   &
  \cellcolor[HTML]{F4C7C3}* &
   &
   &
   \\
\textcite{10.1145/3610073} &
   &
   &
  \cellcolor[HTML]{FCE8B2}* &
   &
   &
   &
   &
   &
   &
   &
  \cellcolor[HTML]{B7E1CD}* &
   \\
\textcite{10.1145/1517744.1517804} &
   &
   &
  \cellcolor[HTML]{FCE8B2}* &
   &
   &
   &
   &
   &
   &
   &
  \cellcolor[HTML]{B7E1CD}* &
   \\
\textcite{10.1145/3064663.3064697} &
   &
   &
   &
  \cellcolor[HTML]{FCE8B2}* &
   &
   &
   &
   &
   &
  \cellcolor[HTML]{B7E1CD}* &
   &
  \cellcolor[HTML]{B7E1CD}* \\
\textcite{10.1145/3460418.3479342} &
   &
   &
   &
  \cellcolor[HTML]{FCE8B2}* &
   &
   &
   &
   &
   &
   &
   &
  \cellcolor[HTML]{B7E1CD}* \\
\textcite{10.1145/3381016} &
   &
   &
   &
  \cellcolor[HTML]{FCE8B2}* &
   &
   &
   &
   &
   &
  \cellcolor[HTML]{B7E1CD}* &
   &
   \\
\textcite{10.1145/3325480.3325491} &
   &
   &
   &
  \cellcolor[HTML]{FCE8B2}* &
   &
   &
   &
   &
   &
   &
  \cellcolor[HTML]{B7E1CD}* &
   \\
\textcite{10.1145/3613904.3642123} &
   &
   &
   &
  \cellcolor[HTML]{FCE8B2}* &
  \cellcolor[HTML]{FCE8B2}* &
   &
  \cellcolor[HTML]{F4C7C3}* &
   &
   &
  \cellcolor[HTML]{B7E1CD}* &
   &
   \\
\textcite{10.1145/3585088.3589381} &
   &
   &
   &
  \cellcolor[HTML]{FCE8B2}* &
   &
  \cellcolor[HTML]{F4C7C3}* &
   &
  \cellcolor[HTML]{F4C7C3}* &
   &
   &
   &
  \cellcolor[HTML]{B7E1CD}* \\
\textcite{10.1145/3643834.3661570} &
   &
   &
   &
  \cellcolor[HTML]{FCE8B2}* &
   &
  \cellcolor[HTML]{F4C7C3}* &
   &
   &
   &
   &
   &
  \cellcolor[HTML]{B7E1CD}* \\
\textcite{10.1145/3316782.3321525} &
   &
   &
   &
  \cellcolor[HTML]{FCE8B2}* &
   &
   &
   &
   &
   &
  \cellcolor[HTML]{B7E1CD}* &
   &
  \cellcolor[HTML]{B7E1CD}* \\
\textcite{10.1145/2660579.2660590} &
   &
   &
   &
  \cellcolor[HTML]{FCE8B2}* &
   &
   &
   &
   &
   &
   &
   &
  \cellcolor[HTML]{B7E1CD}* \\
\textcite{10.1145/3434073.3444669} &
   &
   &
   &
   &
  \cellcolor[HTML]{FCE8B2}* &
  \cellcolor[HTML]{F4C7C3}* &
   &
   &
   &
  \cellcolor[HTML]{B7E1CD}* &
   &
  \cellcolor[HTML]{B7E1CD}* \\
\textcite{10.1145/2544104} &
   &
   &
   &
   &
  \cellcolor[HTML]{FCE8B2}* &
   &
   &
   &
   &
  \cellcolor[HTML]{B7E1CD}* &
  \cellcolor[HTML]{B7E1CD}* &
  \cellcolor[HTML]{B7E1CD}* \\
\textcite{10.1145/3173574.3174069} &
   &
   &
   &
   &
  \cellcolor[HTML]{FCE8B2}* &
  \cellcolor[HTML]{F4C7C3}* &
  \cellcolor[HTML]{F4C7C3}* &
   &
   &
  \cellcolor[HTML]{B7E1CD}* &
   &
   \\
\textcite{10.5555/3523760.3523888} &
   &
   &
   &
   &
   &
   &
   &
   &
   &
  \cellcolor[HTML]{B7E1CD}* &
  \cellcolor[HTML]{B7E1CD}* &
   \\
\textcite{10.1145/3569484} &
   &
   &
   &
   &
  \cellcolor[HTML]{FCE8B2}* &
  \cellcolor[HTML]{F4C7C3}* &
   &
   &
   &
  \cellcolor[HTML]{B7E1CD}* &
   &
   \\
\textcite{10.1145/3329189.3329209} &
   &
   &
   &
   &
   &
  \cellcolor[HTML]{F4C7C3}* &
   &
  \cellcolor[HTML]{F4C7C3}* &
  \cellcolor[HTML]{F4C7C3}* &
   &
   &
  \cellcolor[HTML]{B7E1CD}* \\
\textcite{10.1145/3341163.3346933} &
   &
   &
   &
   &
  \cellcolor[HTML]{FCE8B2}* &
  \cellcolor[HTML]{F4C7C3}* &
   &
   &
   &
  \cellcolor[HTML]{B7E1CD}* &
   &
   \\
\textcite{10.1145/3491102.3502130} &
   &
   &
   &
   &
  \cellcolor[HTML]{FCE8B2}* &
   &
   &
   &
  \cellcolor[HTML]{F4C7C3}* &
  \cellcolor[HTML]{B7E1CD}* &
  \cellcolor[HTML]{B7E1CD}* &
  \cellcolor[HTML]{B7E1CD}* \\ \bottomrule
\end{tabular}%
}
\footnotesize Note: We use abbreviations in the table headings for presentation purposes. Here are the full names of the headings: Tracking->Emotion Tracking and Feedback, Cognitive-> Cognitive Development, Social-> Social Connection, Expression-> Expressive Making, Acceptance-> Self-Acceptance and Compassion; Down-> Down-regulation Goals, Up-> Up-regulation Goals, Non-jud-> Non-judgmental Emotion Support, Ident->Identity Preservation.

\end{table}